\documentclass[fleqn,usenatbib]{mnras}

\usepackage{newtxtext,newtxmath}

\usepackage[T1]{fontenc}

\DeclareRobustCommand{\VAN}[3]{#2}
\let\VANthebibliography\thebibliography
\def\thebibliography{\DeclareRobustCommand{\VAN}[3]{##3}\VANthebibliography}

\usepackage{graphicx}	
\usepackage{amsmath}	
\usepackage{xcolor}

\newcommand\spock{\texttt{SPOCK}\ }
\newcommand\forecaster{\texttt{Forecaster}}
\newcommand\rebound{\texttt{REBOUND}}
\newcommand\mearth{M_{\oplus}}
\newcommand\rearth{R_{\oplus}}
\newcommand\dg{^{\circ}}

\title[Dynamical Packing in the Kepler Systems]{On the Degree of Dynamical Packing in the Kepler Multi-planet Systems}

\author[A. Obertas et al.]{
Alysa Obertas,$^{1,2}$\thanks{E-mail: obertas@astro.utoronto.ca (AO)}
Daniel Tamayo,$^{3}$
Norm Murray$^{2}$
\\
$^{1}$David A. Dunlap Department of Astronomy \& Astrophysics, University of Toronto, 50 St. George Street, Toronto M5S 3H4, Canada\\
$^{2}$Canadian Institute for Theoretical Astrophysics, 60 St. George Street, Toronto M5S 3H8, Canada\\
$^{3}$Department of Physics, Harvey Mudd College, 301 Platt Blvd., Claremont 91711, USA
}

\date{Accepted XXX. Received YYY; in original form ZZZ}

\pubyear{2023}

\begin{document}
\label{firstpage}
\pagerange{\pageref{firstpage}--\pageref{lastpage}}
\maketitle

\begin{abstract}

Current planet formation theories rely on initially compact orbital configurations undergoing a (possibly extended) phase of giant impacts following the dispersal of the dissipative protoplanetary disk. 
The orbital architectures of observed mature exoplanet systems have likely been strongly sculpted by chaotic dynamics, instabilities, and giant impacts.
One possible signature of systems continually reshaped by instabilities and mergers is their dynamical packing. Early Kepler data showed that many multi-planet systems are maximally packed -- placing an additional planet between an observed pair would make the system unstable. 
However, this result relied on placing the inserted planet in the most optimistic configuration for stability (e.g., circular orbits).
While this would be appropriate in an ordered and dissipative picture of planet formation (i.e. planets dampen into their most stable configurations), we argue that this best-case scenario for stability is rarely realized due to the strongly chaotic nature of planet formation.
Consequently, the degree of dynamical packing in multi-planet systems under a realistic formation model is likely significantly higher than previously realized.
We examine the full Kepler multi-planet sample through this new lens, showing that $\sim60-95\%$ of Kepler multi-planet systems are strongly packed and that dynamical packing increases with multiplicity. This may be a signature of dynamical sculpting or of undetected planets, showing that dynamical packing is an important metric that can be incorporated into planet formation modelling or when searching for unseen planets.

\end{abstract}




\section{Introduction}
\label{sec:introduction}

The last decade of exoplanet observations has revolutionized our understanding of planets and planetary systems. Astronomers have a wealth of data on protoplanetary disks \citep[e.g. see][]{and20} and of long-lived multi-planet systems \citep[e.g. see][]{zhu21} which have driven significant innovation and progress in planet formation models \citep[e.g. see][]{dra22}. Most relevant to our investigation is a question of nature versus nurture. Did the Kepler multi-planet systems form in the gas disk with their observed architectures, or did those architectures arise from subsequent dynamical interactions?

Most planet formation theories include a final giant impact phase in which planets excite one another onto crossing orbits and collide (e.g. see \citet{gol04, wya16}, and reviews above). Numerical (i.e. using nbody simulations, e.g. \citet{han13,daw16,mat17,poon20,mac20}) and analytical \citep[e.g][]{tre15} exploration of this phase can approximately reproduce the observed exoplanet population.

In this picture, planetary systems repeatedly destabilize from compact configurations with many planets, leading to gravitational scatterings and planetary mergers that leave behind ever longer-lived, lower-multiplicity architectures \citep{las96}.
This hypothesis, under which orbital architectures are dynamically sculpted throughout planetary systems' lifetimes has been advanced from different perspectives in an exoplanet context by \citet{volk15}, \citet{pu15}, \citet{izi17}.

An extreme, but testable scenario for the outcome of this process is the packed planetary systems (PPS) hypothesis. \citet{bar04b} proposed that these dynamical instabilities result in orbital architectures with no room for additional planets between adjacent pairs, and tested this hypothesis on a sample of observed systems with multiple giant planets \citep[see also][]{bar04a}. The Kepler mission's discovery of hundreds of multi-planet systems allowed for the first test of this hypothesis for lower mass planets, conducted by \citet{fan13}. They found that at least $31-45\%$ of multi-planet systems are dynamically packed, broadly consistent with the PPS hypothesis.

\citet{fan13} obtain lower limits to this dynamical packing by always placing an additional planet in the orbital configuration that one would, on average, expect to be most stable: midway between the observed pair in mutual Hill radii (Appendix~\ref{appendix:mutual-hill-radius}), with all orbits initially circular.
Under this binary definition, the system is dynamically packed only if this favourable configuration for an inserted planet goes unstable within the integration time.
A major advantage to this approach is that it limits the number of integrations to perform (consequently reducing overall computation time).

However, we know that this most optimistic scenario for inserted planets does not hold.
Radial velocity, transit timing, and transit duration measurements have revealed that exoplanet orbits in multi-planet systems typically have small but finite eccentricities \citep[e.g.][]{eyl15,xie16,had17,mil19,eyl19}.
Additionally, even if an equidistant location for an additional planet were stable, there is no guarantee that the planet formation process would create a planet in precisely that location.

Instead, we propose a more general and continuous measure of dynamical packing, which evaluates the stability of a range of configurations for the inserted planet, weighted by the probability that planet formation would create such a configuration.
This more physically meaningful approach to dynamical packing requires a corresponding generative model of planet formation from which to sample configurations for the inserted planet.
Since astronomers are far from such a precise model at the current time, we consider two simple end-member planet-formation scenarios. 

First, for a sufficiently chaotic giant-impact phase, one might expect planet formation to explore the full phase space of possible orbital architectures. 
Not all of these outcomes will be long-term stable, in which case collisions and scatterings yield a new configuration.
By adopting a simplified stability criterion, \cite{tre15} proposed an analytic model in which the chaotic giant impact stage populates space uniformly and planets would only be observed in regions with long-term stability.
This ``ergodic hypothesis'' yields distributions of orbital eccentricities and interplanetary spacings that are broadly consistent with observations \citep{tre15}.
The natural definition of dynamical packing in this framework is then the fraction of phase space between adjacent pairs in which inserting an additional planet leads to instability.
It's important to note that such a strongly chaotic giant impact phase does not necessarily result in disordered systems. \cite{lam23} (see also \citet{gol22}) have shown that numerical integrations of a giant impact phase can nevertheless reproduce the observed preference for similar radii, masses, and spacings within multi-planet systems \citep{wei18, mil17}.

At the opposite extreme, it is possible that dissipation (e.g. from gas or planetesimal disks) plays a strong role in the early stages of planet formation, allowing planetary systems to settle into their most stable configurations and avoid later instabilities.
For example, \citet{ada19, ada20} showed that, under various constraints and simplifying the gravitational interaction between planets, minimization of energy can explain some features of the observed similarities in radii, masses, and spacings within multi-planet systems. If one directly models the interplanetary interactions, dissipative and convergent migration naturally leads to the formation of resonant chains, several of which are now known \citep{mil16, gil17, lug17}, and have presumably not left these resonances since formation \citep{tam17}.
However, the vast majority of multi-planet systems are \textit{not} in such resonant-chain configurations, so if they do form during the gas disk phase, the majority must subsequently break away from their resonances \citep{gol14, izi17, izi18}. The resonance escape may or may not lead to major instabilities.
If it does, the end result of such instabilities and planetary mergers might then be expected to follow the predictions of the ``ergodic hypothesis" mentioned above.
Nevertheless, this ordered framework for planet formation (which we refer to as the ``ordered hypothesis''), provides a useful extreme scenario to compare against.
In this picture, we define an adjacent pair of planets as dynamically packed if there does not exist a long-term stable configuration \textit{anywhere} for an additional planet placed in between the observed pair. 
It is therefore sufficient to examine the stability of the additional planet in its most stable configuration. 
This means that for the ordered hypothesis, dynamical packing reduces to a binary classification similar to that of \citet{bar04b} and \citet{fan13}.

The ergodic definition of dynamical packing requires the exploration of a vast phase space for observed and inserted planets (i.e. planet masses and orbit configurations) and would need a prohibitive number of CPU hours to perform stability assessment using Nbody integrations.
We therefore employ the \spock package, which incorporates analytical dynamical models into flexible machine learning architectures to speed up stability determination by a factor of up to $10^5$ relative to direct N-body integrations \citep{tam20}.

In addition to new computational tools available to perform stability tests, the Kepler mission discovered an order of magnitude more multi-planet systems in the time since \citet{fan13} \citep{tho18}. GAIA DR-2 \citep{gaia18} also allowed for more accurate stellar parameters \citep{ber20b}, and consequently planetary parameters, of the Kepler systems \citep{ber20a}. Furthermore, there have been substantial improvements in observational, theoretical, and modelling work of the mass-radius relationship \citep[e.g.][]{che17}, eccentricities \citep[e.g.][]{eyl15,xie16,had17,mil19,eyl19,he20}, and inclinations \citep[e.g.][]{zhu18a,he20,mil21} of planets in the Kepler sample, all of which strongly impact the dynamics and stability of multi-planet systems.

To build up our results sequentially, we first conduct tests using only the Kepler Quarters 1--6 data in a manner similar to \citet{fan13}. Next, we repeat our analysis with the updated Kepler and GAIA DR-2 combined data. Finally, we move beyond a simple binary classification of dynamical packing under a dissipative, ordered model of planet formation, and evaluate a continuous measure of dynamical packing under a giant impact phase closer to current theories of planet formation. We describe our methods for generating our planet catalogues and performing our dynamical tests in Section~\ref{sec:methods} and present the results of all our tests in Section~\ref{sec:results}. Finally, we summarize our findings and discuss their implications in Section~\ref{sec:conclusions}.

\section{Methods}
\label{sec:methods}

To investigate whether the Kepler multi-planet systems are dynamically packed, we build on \cite{fan13}. We run several sets of stability tests after inserting an additional planet in between adjacent pairs in observed systems. 

\cite{fan13} were limited by the small-number statistics of multi-planet systems detected by the \textit{Kepler} mission in quarters 1--6 (Q1--6).
As a result, they first modelled an underlying multi-planet population and then inserted additional planets into this synthetic population.
With an order-of-magnitude more planet-pairs now discovered, we inserted additional planets directly into observed multi-planet systems from the full Kepler catalogue instead. Additionally, we used improved stellar and planetary parameters informed by the second data release from the GAIA mission \citep{gaia18}.

We build up to our extended analysis in several steps. 
First, for a meaningful comparison to \citet{fan13}, we perform a similar style of tests and analysis as \citet{fan13} using the same observed multi-planet systems in {\it Kepler} data from Q1--6 (described in Sec.~\ref{sec:methods-q16}), except with direct insertion into observed multi-planet systems rather than imposing a particular parameterized distribution to model the underlying planet population (Sec.~\ref{sec:methods-fm13style}).
We then extend the same analysis to the full Kepler-Gaia catalogue (described in Sec.~\ref{sec:methods-keplergaia}).
Finally, we move beyond testing for the stability of a single configuration for an additional planet, to exploring the full phase space of additional-planet configurations (Sec.~\ref{sec:methods-expanded}).

We assessed the stability of planets inserted between observed adjacent pairs using two methods. Our primary tool was \spock~\citep{tam20}, a machine learning code which provides the probability $p_{\spock}$ that a system is stable for $10^9$ orbits. 
Since previous works on this topic express their results in terms of the fraction of systems that are unstable (i.e. dynamically-packed), for consistency we convert this into the probability that a system is unstable within $10^9$ orbits (i.e. $p = 1 - p_{\spock}$). Throughout the remainder of this paper, we use ``probability" to refer to this instability probability. 

\spock~ performs short $10^4$-orbit N-body integrations with \rebound~, and uses these to generate the input features to the machine learning model in order to determine the probability \citep{tam20}. 
This means that \spock~ can check for stability up to $\sim 10^5$ times faster than N-body (N-body becomes proportionally more competitive for configurations that survive less than $10^9$ orbits); across our suite of simulations, \spock~ was about $10^4$ times faster. This allows for a much broader and more rapid phase space exploration than was previously possible. We also verify a subset of our results using slower N-body integrations with the \rebound~ package \citep{rei12}.

\subsection{Kepler Q1--6 Catalogue}
\label{sec:methods-q16}

We used a subset of data from Kepler Quarters 1--6\citep{bat13}, as these were the data used for the analysis in \citet{fan13}, using these stellar and planetary properties to generate our catalogue of observed multi-planet systems (which we refer to as the ``Q1--6 catalogue"). We used this catalogue for a comparison of our methods with \citet{fan13}, so we adopted their same cuts, retaining only planets that satisfied all of:

\begin{eqnarray}
    P \leq 200 \mathrm{days} \nonumber \\
    1.5 \rearth \leq R \leq 30 \rearth \nonumber \\
    \mathrm{SNR} \geq 10 \nonumber
\end{eqnarray}

Where $P$ is the period of the planet's orbit, $\rearth$ is the Earth's radius, $R$ is the planet's radius, and the signal-to-noise ratio is denoted as SNR. Some of these cuts removed planets in multi-planet observed systems, leaving only a single planet. These observed systems were dropped from our catalogue. Our Q1--6 catalogue contains 140 planets around 60 stars. There are 43 observed systems with $N=2$ planets, 14 with $N=3$, and 3 with $N=4$. Note that these are observed multiplicities after making our cuts.

\subsection{Kepler-GAIA DR2 Catalogue}
\label{sec:methods-keplergaia}

To generate our up-to-date catalogue of multi-planet systems (which we refer to as our ``Kepler-GAIA catalogue"), we first started by generating a catalogue containing a subset of the planets from the Kepler Cumulative List on the NASA Exoplanet Archive\citep{cumulative}, filtering to include only planets with a disposition of ``CANDIDATE"\footnote{Note: the ``Disposition Using Kepler Data" column does not have ``CONFIRMED" as a possible value.} and a radius of ``not NULL". This was then merged with the more accurate stellar masses from \citet{ber20a} and planet radii from \citet{ber20b}, as informed by GAIA DR-2, only retaining stars and planets present in all three of the Kepler Cumulative List, \citet{ber20a}, and \citet{ber20b}. We note that \citet{ber20b} explicitly excluded exoplanets whose host stars likely have a binary companion(s)\footnote{Specifically, stars with a renormalised unit-weight error (RUWE) $> 1.2$.}

We then kept only planets with physically plausible radii (the fraction of ``CANDIDATE" planets with $R>20 R_\oplus$ is negligible), and fractional errors in the planet radius smaller than unity:

\begin{eqnarray}
   R < 20 \rearth \nonumber \\
   \sigma_{R,+} / R < 1 \nonumber \\
   \sigma_{R,-} / R < 1. \nonumber
\end{eqnarray}

After making these cuts, we kept only planets in multi-planet systems. Additionally, we removed the two planets around KIC 3245969, as their periods differ by approximately 7 seconds. Our Kepler-GAIA catalogue contains 1536 planets around 617 stars. There are 408 systems with $N=2$, 141 with $N=3$, 48 with $N=4$, and 20 with $N\geq5$. Like the Q1--6 catalogue, these are the observed multiplicities after making our cuts.

A system's observed multiplicity is not necessarily its intrinsic multiplicity. Additional planets have been found in Kepler systems through methods other than the survey's detection pipeline, such as the non-transiting planet Kepler-20 g \citep{buc16} (although its radial velocity signal may be caused instead by stellar activity \citep{nav20}) or the low signal-to-noise planets Kepler-80 g and Kepler-90 i \citep{shal18}. These planets are not included in our study as they are not present in the cumulative Kepler catalogue, although we note that this would negatively impact the stability of inserted planets in our tests.

It is likely that there are also undetected planets around the stars in our Kepler-GAIA catalogue, although the quantity is sensitive to planet inclinations (e.g. as first explored in the Kepler multi-planet systems by \citet{lis11}). Estimates of the average number of planets within $\leq400$ days around planet-hosting stars in the Kepler sample generally range from $\sim3-6$ \citep{tra16, zhu18a, zin19, san19}, which means that we would expect an additional $\sim300-2150$ ($\sim20-140\%$) undetected planets in our Kepler-GAIA catalogue (with the caveat that we have excluded stars with only a single transiting planet). This could have an impact on the results of our stability tests if undetected planets are typically between adjacent observed pairs (e.g. if they are non-transiting like Kepler-20g, but don't produce strong radial velocity signals).

Studies examining intrinsic multiplicity don't comment on the locations of undetected planets, so we conducted a simple test to estimate the frequency that an observed (i.e. transiting) adjacent pair has at least one undetected planet in between them. We found that $\approx 24\%$ of pairs have at least one in-between undetected planet (see Appendix \ref{appendix:undetected-planets} for more details). Even though this is a non-negligible proportion of pairs, we opted to proceed with our stability tests as if all observed pairs in our Kepler-GAIA catalogue are truly adjacent (rather than e.g. attempt to account for or incorporate the possible presence of in-between undetected planets). In-between undetected planets can only make a system \textit{more} dynamically packed, which means our analysis will provide robust lower limits on packing.

Planets at longer periods have been observed in many Kepler systems (e.g. \citet{sch14,wan15}), meaning that some of the systems in our catalogue may also have undetected planets at larger periods affecting the stability of the inserted planets in our tests. Indeed, several studies (e.g. \citet{zhu18b,bry19,her19,ros22}) have noted that cold giants typically have a compact inner set of planets. Given these studies, \citet{zhu21} estimate that the probability of a cold Jupiter given a set of inner super Earths is $P(CJ|SE) \approx 30\%$; however this relationship was recently questioned by \citet{bon23}, who did not find evidence supporting the relationship. If outer giant planets are present, \cite{mil22} recently argued that these undetected planets must typically lie significantly farther out, which means they likely would not be the predominant source of dynamical perturbations for the planets we insert between adjacent pairs in our Kepler-GAIA catalogue. In any case, their presence would only render each system \textit{more} unstable, similar to additional nearby planets. Similar to in-between undetected planets, this means our analysis will give lower limits on dynamical packing.

\subsection{Fang \& Margot Style Instability Tests}
\label{sec:methods-fm13style}

The goal of \citet{fan13} was to test the packed planetary systems hypothesis \citep{bar04b} by inserting an additional planet in between adjacent pairs. The eccentricities and positions of inserted planets were chosen to maximize stability. If, under those conditions, a system could not host an additional planet, then they considered it to be dynamically packed.

Our first tests were in a similar style as \citet{fan13} (which we refer to as ``FM13-style" tests) where we inserted a planet such that it has equal semimajor-axis separations from each neighbour, in units of the mutual Hill radius (see Appendix~\ref{appendix:mutual-hill-radius}).
This choice tries to account for situations where the observed planets have very different masses, so one would expect the most stable position for the inserted planet to be farther away from the massive body.
We note that while this placement should be beneficial for stability on average, it could also sometimes correspond to one of many strongly chaotic regions (e.g. near the edge of a strong mean motion resonance \citep{obe17}). 

Additionally, as a best-case scenario for stability (and without better information at hand), \cite{fan13} assumed all orbits were circular. 
We now know from radial velocities \citep{mil19}, transit duration variations \citep{eyl15,xie16,eyl19}, and transit timing variations \citep{had17}, that Kepler systems exhibit small but finite orbital eccentricities, which strongly (and typically negatively) affect stability \citep{had18, tam21b, yee21}. Regardless, we retain these two choices for our FM13-style tests for the sake of comparison and leave the exploration of different placements and non-zero eccentricities for our expanded parameter space tests (see \ref{sec:methods-expanded} below).

We ran these FM13-style tests with both the Q1--6 and the Kepler-GAIA catalogues. Although we expect poor statistics for the FM13-style tests on the Q1--6 catalogue, this allows us to compare catalogues using a similar methodology.

For these tests, we created a \rebound~simulation with the star (using the mass in its respective catalogue). 
For the planets' orbits, following \citet{fan13}, we adopted an eccentricity of zero and an inclination drawn from a Rayleigh of Rayleigh distribution\footnote{This is a Rayleigh distribution where the Rayleigh parameter itself is also a Rayleigh distribution.} with parameter $\sigma_{\sigma_i} = 1\dg$. We adopt a uniformly drawn longitude of ascending node ($\Omega$), argument of pericentre ($\omega$), and true anomaly ($f$). 
The observed planets are assigned their measured orbital periods, while the inserted planet is placed equidistant between them (in mutual Hill radii). 

The masses of the observed planets in our two catalogues were calculated in two different ways based on their observed radii. For the Q1--6 catalogue, we used the same mass-radius relationship as \citet{fan12} and \citet{fan13} (note that our expression below is given in terms of $\rearth$ and $\mearth$ rather than $R_{\mathrm{Jupiter}}$ and $M_{\mathrm{Jupiter}}$).

 \begin{eqnarray}
    \log_{10}\left({\frac{M}{\mearth}}\right) = 0.215689\left(\frac{R}{\rearth}\right) + 0.2412,~\mathrm{for}~R < 11.6595 \rearth \label{eq:mass-radius} \\
    \log_{10}\left({\frac{M}{\mearth}}\right) = -0.0448137\left(\frac{R}{\rearth}\right) + 3.279,~\mathrm{for}~R \geq 11.6595 \rearth
 \end{eqnarray}

For the Kepler-GAIA catalogue, we generated a sample of radii for each planet by using a two-sided normal distribution with mean value $R$ and standard deviations $\sigma_{R,+}$ and $\sigma_{R,-}$. Then, we calculated mass distributions for each planet using \forecaster ~\citep{che17}. Finally, we used the median mass of these distributions as each planet's mass.

Following \cite{fan13}, the mass of the inserted planet was determined by using eq.~\ref{eq:mass-radius} for a planet with $R=1.5\rearth$, resulting in $M = 3.671\mearth$. This mass was adopted for the inserted planet in our FM13-style tests for both the Q1--6 and Kepler-GAIA catalogues.

For each system in the catalogue with an observed planet multiplicity $N$, we generated $N-1$ \rebound~simulations, each with only a single additional inserted planet. For example, an $N=3$ system would have two simulations: one with a planet inserted between the inner and middle planets and a second with a planet inserted between the middle and outer planets. For the Q1--6 catalogue, we had 43 pairs belonging to a system with an observed multiplicity of $N=2$, 28 with $N=3$, and 9 with $N=4$. For the Kepler-GAIA catalogue, we had 408 pairs belonging to a system with an observed multiplicity of $N=2$, 282 with $N=3$, 144 with $N=4$, and 85 with $N\geq5$. Note the numbers of pairs differ from the number of systems (Sec.~\ref{sec:methods-q16}--\ref{sec:methods-keplergaia}) if $N\geq3$.

We assessed the stability of planets inserted between observed adjacent pairs using Nbody integrations and \spock. For the Nbody integrations, we used \rebound~with \texttt{WHFAST} \citep{rei15} and a timestep of 5\% of the inner planet's initial period. The \rebound~simulation was integrated for the maximum integration time (see below), until a close encounter occurred, or until a planet escaped. We set each planet's radius in the \rebound~simulation as its Hill sphere (i.e. $R=a(M/3M_{\star})^{1/3}$) and used a collision as our criterion for a close encounter. We set a maximum distance for the \rebound~simulation as 100 times the outer planet's initial distance and used this as our criterion for an escape. For the Q1--6 catalogue, we ran one set of integrations with a maximum integration time of $10^9$ orbits of the innermost planet (using its initial period). For the Kepler-GAIA catalogue, we ran two sets of integrations: one with a maximum integration time of $10^9$ orbits and a second with a maximum integration time of $10^8$ years.

We used \spock~to obtain the probabilities of configurations being unstable within $10^9$ orbits for a planet inserted between an adjacent pair for all sets of pairs in both catalogues. Conveniently, \spock~calculates its probabilities by being given a \rebound~simulation. This allowed us to pass to \spock~the exact same initial \rebound~simulations that were used for our Nbody integrations.
We note that because the dynamics are chaotic, N-body integrations run on two sets of initial conditions which are separated by machine precision will yield two different but equally valid instability times.
Nevertheless, \cite{hus20} find that the width of such instability time distributions is typically small compared to the mean instability time, so this is only a quantitative concern for the fraction of configurations with mean instability times within approximately one dex of the integration time (e.g., within $\sim10^{8.5}-10^{9.5}$ orbits for our $10^9$ orbit runs).
This therefore does not significantly impact our results.
In Appendix \ref{appendix:nbody-spock comparison} we show that \spock compares well against N-body integrations for a representative subset of our suites of simulations.

\subsection{Expanded Parameter Space Instability Tests}
\label{sec:methods-expanded}

The significant computational savings of \spock vs.~direct N-body integrations allows us to efficiently explore a broad parameter space of orbital configurations for inserted planets.
Rather than testing a single configuration (equidistant in Hill radii, circular orbits) for an additional planet between each observed pair \citep{fan13}, we drew 10 000 orbital configurations randomly sampling planets' physical and orbital parameters.

Specifically, we initialized a star with its Kepler-GAIA catalogue mass. 
The transiting (observed) planets were assigned their Kepler-GAIA catalogue periods. 
To initialize their masses, we began by drawing a radius from a two-sided normal distribution (with the catalogue values of $R$, $\sigma_{R,+}$, and $\sigma_{R,-}$ as the mean and standard deviations).
We then passed this radius to the \forecaster~package \citep{che17}, which uses it to estimate a probability distribution function (pdf) for the mass, and returns a random sample from this pdf.

While some of the planets in our Kepler-GAIA catalogue have measured masses from radial velocity (RV) or transit timing variation observations, our ensemble results are not sensitive (within error) to the use of measured versus \forecaster~observed planet masses. We chose to use \forecaster~to generate individual mass pdfs for all planets in the Kepler-GAIA catalogue in order to apply a consistent methodology for generating our \rebound~simulations. To check if our results were sensitive to the choice of mass for observed planets, we compared our two summary metrics (see Sec.~\ref{sec:results-ergodic} \& Table~\ref{tab:summary-expanded-mean} and~\ref{sec:results-ordered} \& Table~\ref{tab:summary-expanded-msc}) for a subset of planets that are present both in Table~10 of \citet{bon23} and in our Kepler-GAIA catalogue. Although the RV masses from \citet{bon23} are typically higher than the median \forecaster~mass (which would only make systems \textit{more} unstable), our summary metrics typically differed well within our estimated errors. Using the median \forecaster~mass, the 4 observed adjacent pairs (8 planets) in 4 systems with measured RV masses have mean ($\mu$) and standard deviation ($\sigma$) values of the RV-\forecaster~masses ratio (i.e. $M_{RV} / M_f$) of $\mu \sim 1.3$ and $\sigma \sim 0.5$\footnote{For all 14 planets with measured RV masses (but not necessarily adjacent pairs), $\mu \sim 1.5$ and $\sigma \sim 0.5$.}, with both summary metrics typically having a small difference (i.e. slightly less stable) within our errors (see Tables~\ref{tab:summary-expanded-mean} and~\ref{tab:summary-expanded-msc} plus the discussion in Appendix~\ref{appendix:msc} for our error estimates). Expanding to the set of 28 planets (18 adjacent pairs) in 9 multi-planet systems with measured or constrained (i.e. $M_{RV} < $ some value) masses, both summary metrics similarly differed within our errors despite much larger mean and standard deviation values of the masses ratio ($\mu \sim 2.7$, $\sigma \sim 2.6$)\footnote{Exclusive of Kepler-37 b, which has $R = 0.28${\raisebox{0.5ex}{\tiny$^{+0.03}_{-0.02}$}}$\rearth$ in our Kepler-GAIA catalogue but $M_{RV} / M_f < 75.7$. Including it gives $\mu \sim 5.1$, $\sigma \sim 13.2$.}.

\begin{figure}
  \centering
    \includegraphics[width=\columnwidth]{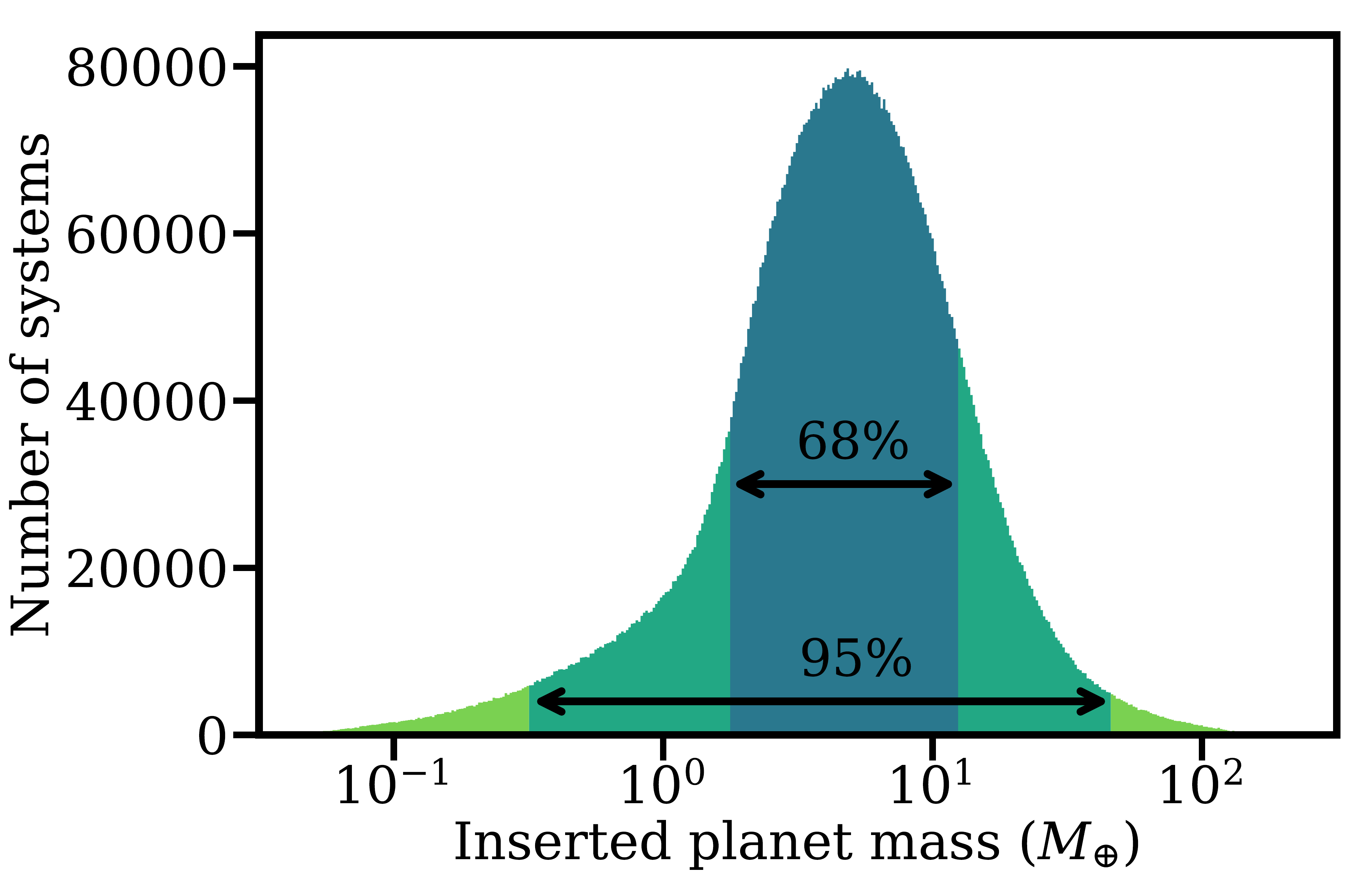}
    \caption{Histogram of the inserted planet masses for all 9 190 000 sets of adjacent pairs examined. We obtained masses by applying \forecaster~\citep{che17} to a randomly drawn radius from the Kepler-GAIA catalogue. The median mass is $4.84\mearth$, 68\% are within $1.75\mearth \leq M \leq 12.56\mearth$, and 95\% of are within $0.32\mearth \leq M \leq 46.41\mearth$. 0.47\% of masses are below $0.1\mearth$ and 1.45\% of masses are above $100\mearth$.}
    \label{fig:inserted-planet-mass-hist}
\end{figure}

For the hypothetical inserted planet, we sampled the orbital period uniformly between the two observed periods.
We then randomly drew a physical radius from the observed Kepler-GAIA catalogue, and sampled a corresponding mass as described above with the \forecaster~package \citep{che17}. 
The distribution of masses of all inserted planets is shown in Figure~\ref{fig:inserted-planet-mass-hist}. The median mass is $4.84\mearth$, and 95\% have $0.32\mearth \leq M \leq 46.41\mearth$.\footnote{Our FM13-style tests instead use $M=3.671\mearth$, which corresponds to the smallest radius considered by \cite{fan13} of $1.5\rearth$. This falls within the central 68\% interval of our adopted mass distribution.} 
We note that our results do not strongly depend on this choice for how the mass of the inserted planet was sampled (see Fig.~\ref{fig:stacked-grid-plots}).

Finally, all eccentricities were drawn from a Rayleigh distribution with parameter $\sigma_e$; inclinations were drawn from a Rayleigh distribution with parameter $\sigma_i$; and orbital angles $\Omega$, $\omega$, and $f$ were drawn uniformly.

Our eccentricity and inclination distributions were motivated both by observational work to infer their values using methods such as radial velocities, transit duration variations, and transit timing variations \citep[e.g.][]{xie16, had17, zhu18a, eyl19, mil19, mil21}. Additionally, small inclinations have a small impact on the stability of transiting planets \citep{tam21b}, so we use $\sigma_e\sim\sigma_i$ for simplicity. To represent a "low end" and "high end" of the range from observations, we chose two sets of values for $(\sigma_e, \sigma_i)$: $(0.01, 0.5\dg)$ and $(0.05, 2.5\dg)$.

We note that while we sample periods uniformly, we do not do the same in the eccentricity degrees of freedom, which would correspond to sampling uniformly in $e^2$ \citep{tre15} (i.e. this would be an ergodic sampling). Uniform sampling favours the largest eccentricities, while our choice of Rayleigh distributions favours particular mean eccentricity values that can be calibrated against observations. An ergodic sampling of eccentricities would thus only increase our quoted estimates of dynamical packing.

As in our FM-13 style tests, for a system with $N$ observed planets, we tested its stability $N-1$ times with the additional planet inserted between the $N-1$ sets of observed adjacent pairs. We thus similarly test the dynamical packing of 919 observed, adjacent planet pairs with the same multiplicity breakdown listed in the previous subsection. 
However, instead of testing a single configuration for the inserted planet, we generated 5000 configurations (not just sampling parameters of the inserted planet, but also of all of the system's observed planets) for each of the 919 sets of adjacent pairs using the two sets of values for $(\sigma_e, \sigma_i)$, giving a total of 9 190 000 tested configurations.

\section{Results}
\label{sec:results}

Our ultimate goal is to investigate the dynamical packing of the Kepler multi-planet systems in the Kepler-GAIA catalogue using our expanded definition of dynamical packing. Before conducting those tests, we performed a series of simplified tests to serve as stepping stones. This allowed us to build up to our expanded tests, but also to provide context and to see how our results compare to those of \citet{fan13}'s previous study on dynamical packing in Kepler multi-planet systems. 
Since we are concerned with the instability (and therefore the dynamical compactness) of the ensemble of multi-planet systems rather than the instability of each individual original system, we focus on the overall results of our tests and report summary metrics of system instability from our tests.

We first present the results of our FM13-style tests on the Kepler Q1--6 catalogue (the data used by \citet{fan13}), followed by our FM13-style tests on the Kepler-GAIA catalogue (which contains an order of magnitude more planets than the Q1--6 catalogue). Finally, we present the results of our expanded parameter space tests with a continuous definition of dynamical packing.

\subsection{FM13-Style Tests for Kepler Q1--6 data}
\label{sec:results-q16}

First, we test for stability when inserting a planet equidistant (in mutual Hill radii) between adjacent pairs, which we refer to as ``FM13-style tests". Our methodology is similar to \citet{fan13}, except that we inserted the additional planets into observed multi-planet systems rather than into generated Kepler-like systems. 

\begin{table}
\centering
\caption{Proportions of unstable configurations for the FM13-style tests applied to the Kepler Q1--6 catalogue. Column 1 shows the host system's observed planet multiplicity, with the number of analysed adjacent pairs in parentheses. Column 2 shows the percentage of Nbody integrations that went unstable within $10^9$ orbits ($\pm$ binomial error; $\sqrt{p(1-p)/n_{\mathrm{tot}}}$). Column 3 shows the mean probability of going unstable within $10^9$ orbits ($\pm$ standard error of the mean). Column 4 is from Table~1 of \citet{fan13}.}
\label{tab:summary-fm13-style-q16}
\begin{tabular}{cccc}
\hline
Multiplicity & \multicolumn{2}{c}{Kepler Q1--6} & FM13 \\
& $10^9$ orbits & \spock & \\
\hline
N = 2 (43) & 23.3 $\pm$ 6.4\% & 30.4 $\pm$ 5.1\% & $\geq$31\% \\
N = 3 (28) & 46.4 $\pm$ 9.4\% & 47.8 $\pm$ 7.4\% & $\geq$35\% \\ 
N = 4 (9) & 66.7 $\pm$ 15.7\% & 62.1 $\pm$ 13.5\% & $\geq$45\% \\
\hline
\end{tabular}
\end{table}

Table~\ref{tab:summary-fm13-style-q16} shows the results for our FM13-style tests applied to adjacent pairs in the Kepler Q1--6 catalogue, broken down by the observed multiplicities of the pairs' host systems (shown in column 1). The second column shows the percentage of configurations that went unstable in N-body integrations of $10^9$ orbits and the estimated statistical error.

The numbers in the third column are calculated differently. SPOCK does not yield a binary result of stable or unstable, but rather a probability in the range [0, 1]. The expected total number of unstable systems is then simply the sum of those probabilities \citep[see][for more discussion]{tam21a}. The third column therefore lists the estimated fraction of unstable systems, which is equivalent to the mean SPOCK probability that systems go unstable within $10^9$ orbits, and the estimated statistical error.

The fourth column shows the corresponding values from Table 1 of \citet{fan13} (which does not include error estimates), where they drew synthetic systems from their modelled planet population, and performed N-body integrations for $10^8$ {\it years}. For many of the Kepler systems, $10^9$ orbits is substantially less than $10^8$ years. 
We show below that this discrepancy does not affect our conclusions.

We see that the fraction of unstable systems from our Nbody integrations is consistent with the mean SPOCK probability for all multiplicities, within our statistical errors (specified in the Table caption). 
Given the small number of multi-planet systems in the Kepler Q1--6 catalogue and the resulting large error bars, our results are also broadly consistent with \citet{fan13}. 

While \citet{fan13} do not comment on it (presumably because of the large statistical errors), these results all weakly suggest that the proportion of dynamically packed systems may increase with planet multiplicity.

\subsection{FM13-Style Tests for the full Kepler-GAIA catalogue} \label{sec:results-keplergaia}

\begin{table}
\centering
\caption{Proportions of unstable configurations for the FM13-style tests applied to the Kepler-GAIA. Column 1 shows the host system's observed planet multiplicity, with the number of analysed adjacent pairs in parentheses. Columns 2 and 3 show the percentage of Nbody integrations that went unstable within $10^8$ years and $10^9$ orbits, respectively. Column 4 shows the mean probability of going unstable within $10^9$ orbits. Column 5 is from Table~1 of \citet{fan13}. Errors are calculated as in Table \ref{tab:summary-fm13-style-q16}.}
\label{tab:summary-fm13-style-keplergaia}
\begin{tabular}{ccccc}
\hline
Multiplicity & \multicolumn{3}{c}{Kepler-GAIA Catalogue} & FM13 \\
& $10^8$ years & $10^9$ orbits & \spock & \\
\hline
N = 2 (408) & 19.6 $\pm$ 2.0\% & 16.7 $\pm$ 1.8\% & 26.1 $\pm$ 1.6\% & $\geq$31\% \\
N = 3 (282) & 33.7 $\pm$ 2.8\% & 29.8 $\pm$ 2.7\% & 32.3 $\pm$ 2.0\% & $\geq$35\% \\ 
N = 4 (144) & 46.5 $\pm$ 4.2\% & 40.3 $\pm$ 4.1\% & 38.8 $\pm$ 2.9\% & $\geq$45\% \\
N$\geq$ 5 (85) & 65.9 $\pm$ 5.1\% & 60.0 $\pm$ 5.3\% & 50.1 $\pm$ 4.1\% & \\
\hline
\end{tabular}
\end{table}

The full Kepler-GAIA catalogue provides an opportunity to test this trend with multiplicity on an order-of-magnitude larger sample, which we present in Table~\ref{tab:summary-fm13-style-keplergaia}. 
The columns are analogous to Table~\ref{tab:summary-fm13-style-q16}, with the addition of column 2 which shows the percentage of configurations that went unstable in Nbody integrations of $10^8$ years. The fifth column again lists the result from \citet{fan13}.

With a significantly larger observed planet sample, these tests confirm the trend of an increasing proportion of dynamically packed systems with increasing multiplicity. 

With improved planet statistics compared to Q1--6 and correspondingly smaller error bars, we can see some discrepancies between N-body and SPOCK (specifically for $N=2$). If a sample predominantly consists of stable configurations, \spock will overestimate the sample's mean probability (see Appendix~\ref{appendix:nbody-spock comparison}).

Although our expanded parameter space tests (discussed below in Sec.~\ref{sec:results-expanded}) only have mean probabilities using \spock, this discrepancy is not an issue. Since dynamical packing increases with multiplicity, an overestimation of a highly-stable sample's mean probability (i.e. for lower multiplicities) would only strengthen this trend. Additionally, relaxing our assumption of circular orbits varies the dynamical packing fractions of tens of percent (e.g. Table~\ref{tab:summary-expanded-mean}) and our systematic errors in determining the most stable configuration are on the order of ten percent (Appendix~\ref{appendix:msc}).
Thus, the discrepancies between N-body and SPOCK for $N=2$ are not a strong factor limiting our investigation, and the speed increase provided by SPOCK will allow us to perform much more expansive phase space investigations. 

Note that SPOCK yields probabilities of stability over $10^9$ orbits, while \cite{fan13} performed N-body integrations over $10^8$ years. In the Kepler-GAIA catalogue, $10^9$ orbits correspond to a timescale that is less than $10^8$ years for nearly all systems (the median time corresponding to $10^9$ orbits is $1.52 \times 10^7$ years). As a check, we also perform $10^8$-year N-body integrations and list the results in the second column. 
As expected, a higher proportion of systems are unstable when integrated for $10^8$ years compared to $10^9$ orbits. 
However, given that the distribution of instability times in compact planetary systems is roughly log-uniform \citep{volk15}, there is not a substantial difference between the number of unstable systems in our two tests. 

\subsection{Expanded Parameter Space Tests}
\label{sec:results-expanded}

Instead of testing only a single point estimate for the most stable configuration of a planet inserted between each observed adjacent pair as in our FM13-style tests, our expanded parameter space tests use \spock to explore the vast parameter space available to the inserted planet. We generate 10000 configurations of observed systems with an inserted planet between each adjacent pair tested, drawing orbital periods for the inserted planet uniformly within the gap, and drawing the eccentricities and inclinations from Rayleigh distributions with parameters $\sigma_e$ and $\sigma_i$ (5000 configurations for two sets of $\sigma_e$,$\sigma_i$ values), respectively. 

\begin{figure*}
  \centering
    \includegraphics[width=0.8\textwidth]{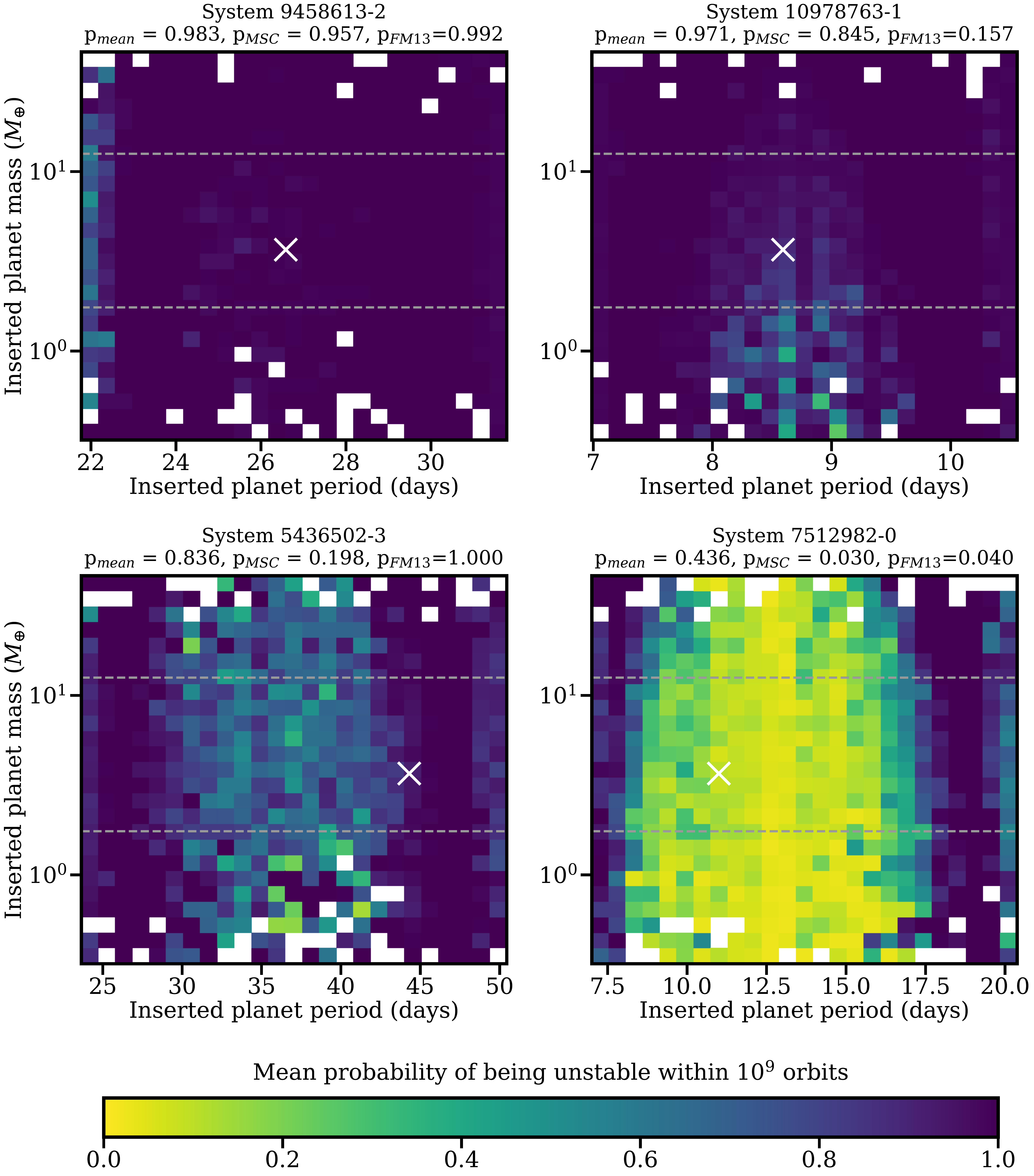}
    \caption{Example probability distributions of being unstable within $10^9$ orbits for four different sets of adjacent pairs. An additional planet was inserted between the adjacent pair and the system was sampled 5000 times (using Rayleigh distributions with $\sigma_e=0.01$, $\sigma_i=0.5^{\circ}$). Each panel is a 2D histogram on a 25 by 25 grid showing the mean probability of all sampled configurations with an inserted planet whose mass and period fall within the bounds of each grid cell (white spaces have no inserted planets in that region). The white X shows the inserted planet's mass and period for the adjacent pair's FM13-style test. Each panel's title shows its KIC, the placement of the inserted planet (0 is between the innermost planet and the second planet), the mean probability ($\mathrm{p}_{mean}$), the most stable configuration probability ($\mathrm{p}_{MSC}$), and the probability of the pair's FM13-style test ($\mathrm{p}_{FM13}$). Inserted planet periods are distributed uniformly in period and the masses according to Figure~\ref{fig:inserted-planet-mass-hist} (the vertical limits correspond to 95\% of masses and 68\% of masses fall within the region marked by the horizontal dashed lines).}
    \label{fig:sample-grid-plots}
\end{figure*}

Fig.~\ref{fig:sample-grid-plots} shows examples of the probability distributions for four different adjacent pairs in our expanded parameter space test using $\sigma_e=0.01$ and $\sigma_i=0.5\dg$. 
Each panel shows the probability distribution projected onto a 25 by 25 grid of the inserted planet's mass vs. its orbital period.
The colour shows the mean probability of being unstable within $10^9$ orbits, marginalizing over all the parameters not visible in this projection (i.e., the eccentricities, inclinations, and orbital angles for all planets, and the masses sampled within observational uncertainties for the two detected planets on either side, see Sec.~\ref{sec:methods-expanded}).
Note that the periods are distributed uniformly, whereas the masses are distributed according to Fig.~\ref{fig:inserted-planet-mass-hist} (i.e. the number density of configurations within each grid cell is uniform horizontally but \textit{not} uniform vertically). The vertical limits correspond to 95\% of the mass distribution and the horizontal dashed lines encompass 68\% of the mass distribution\footnote{The vertical extents of each panel in Figs.~\ref{fig:sample-grid-plots} and~\ref{fig:stacked-grid-plots} are limited to 95\% of masses, but no cuts were made to compute values in Tables~\ref{tab:summary-expanded-mean} and~\ref{tab:summary-expanded-msc}.}.

Each panel in Fig.~\ref{fig:sample-grid-plots} is labelled with the KIC of the adjacent pair's host star and the location of the pair within that system (with 0 meaning that the additional planet is inserted in between the innermost observed planet and its closest neighbour, 1 corresponding to between the second and third, etc.). 

For some adjacent pairs, the first and last columns on some of the grids show higher probabilities of stability (visible in both Figures~\ref{fig:sample-grid-plots} and~\ref{fig:stacked-grid-plots}). This corresponds to co-orbital configurations of the inserted planet and either the inner or outer observed planet. 
Although these particular \spock probabilities are not particularly reliable (see Appendix~\ref{appendix:co-orbitals}), the configurations represent only a small proportion of the total parameter space. We computed quantities presented in this paper including and excluding these columns and they all differed within error (with the exception of Table~\ref{tab:summary-expanded-msc-minimum}). For simplicity, we therefore present our results without making any period cuts for the inserted planet.

The systems in Figure~\ref{fig:sample-grid-plots} highlight four qualitatively different scenarios.
The white X in each panel marks the location for the additional planet using the procedure of \citet{fan13} (Sec.~\ref{sec:methods-fm13style}), with corresponding probability $p_{FM13}$ of being unstable.
As discussed in Sec.~\ref{sec:introduction}, one might expect an ordered, dissipative planet formation process to yield orbital configurations for additional planets that promote stability, like those adopted by \citet{fan13} (equal Hill-radius spacing, circular orbits).
The relevant summary metric for stability under this hypothesis is then (Sec.~\ref{sec:results-ordered}) the lowest probability of instability found in our grid, i.e., the Most Stable Configuration (MSC), with corresponding probability $p_{MSC}$.
In a strongly chaotic, ``ergodic" picture of planet formation where all grid points have uniform probability, the relevant measure of dynamical packing is the mean probability of instability across the grid $p_{mean}$ (Sec.~\ref{sec:results-ergodic}).

While both the FM13-style and the MSC optimize for stability, two competing effects cause deviations between $p_{FM13}$ and $p_{MSC}$: relaxing the assumption of circular orbits of \citet{fan13} should drive up the fraction of dynamically packed systems, whereas exploring a broader parameter space for inserted planets should allow us to find more stable configurations and drive down the fraction of dynamically packed systems. 

The top left and bottom right panels in Fig.~\ref{fig:sample-grid-plots} show cases where the \citet{fan13} placement roughly matches the MSC. The top left panel is a case where essentially the whole region is unstable, while the bottom right panel shows a case where the system remains stable even when a planet is inserted at almost any period and a broad range of masses. Although observed adjacent pairs typically do not have undetected planets located in between them (see Appendix~\ref{appendix:undetected-planets} for our estimated frequency), pairs with probability grid plots visually similar to the lower right panel (i.e. they have smaller values of $p_{mean}$ and look very yellow) are the most suitable candidates to find this type of unseen planet.

The remaining two panels of Fig.~\ref{fig:sample-grid-plots} show cases where the simple method of \citet{fan13} gives a poor estimate of the stability of the system.
The bottom left shows an example with a much more massive inner observed planet, so that the equal-Hill-radii FM13-style placement puts the inserted planet much closer to the low-mass outer neighbor. 
We see that in this case, that placement (probability of being unstable $p_{FM13} = 1$) puts the inserted planet outside the broad stable region ($p_{MSC} = 0.198$). 

The top right panel shows a system that has high probability of being unstable in the expanded parameter space test ($p_{MSC} = 0.845$) but low probability in the FM13-style test ($p_{FM13} = 0.157$).
This is because the FM13-style placement is located near the bounds of a region with \textit{slightly} higher stability on average and happens to be one such configuration that is more stable in that region (i.e. it's an outlier).

\subsubsection{Dynamical Packing Under the Ergodic Hypothesis of Planet Formation}
\label{sec:results-ergodic}

\begin{table}
\centering
\caption{Proportions of unstable configurations for the expanded parameter space tests applied to the Kepler-GAIA catalogue. Column 1 shows the host system's observed planet multiplicity, with the number of analysed adjacent pairs in parentheses. Columns 2 and 3 show the mean probability (using each pair's \textbf{mean} probability, i.e. the proportion of parameter space which is unstable) for the tests with ($\sigma_e, \sigma_i$) = ($0.01, 0.5\dg$) and ($0.05, 2.5\dg$), respectively. Column 5 is from Table~1 of \citet{fan13}. Errors are calculated as in Table \ref{tab:summary-fm13-style-q16}.}
\label{tab:summary-expanded-mean}
\begin{tabular}{cccc}
\hline
Multiplicity & $\sigma_e=0.01$, $\sigma_i=0.5\dg$ & $\sigma_e=0.05$, $\sigma_i=2.5\dg$ & FM13 \\
\hline
N = 2 (408) & 58.9 $\pm$ 1.3\% & 77.7 $\pm$ 1.2\% & $\geq$31\% \\ 
N = 3 (282) & 67.8 $\pm$ 1.4\% & 85.5 $\pm$ 1.2\% & $\geq$35\% \\
N = 4 (144) & 74.8 $\pm$ 1.6\% & 92.9 $\pm$ 0.9\% & $\geq$45\% \\
N$\geq$ 5 (85) & 81.4 $\pm$ 1.9\% &  95.6 $\pm$ 1.0\% & \\
\hline
\end{tabular}
\end{table}

The ergodic hypothesis posits that planet formation is sufficiently chaotic to yield orbital configurations that roughly fill phase space uniformly. Accordingly, we assess dynamical packing in Table~\ref{tab:summary-expanded-mean} using the mean probability $p_{mean}$ for each adjacent pair's probability distribution of being unstable within $10^9$ orbits (i.e. the estimated proportion of our sampled space which is unstable for an inserted planet).

The two central columns explore the effects of finite eccentricities and inclinations, with representative low ($\sigma_e = 0.01$, $\sigma_i=0.5\dg$) and high ($\sigma_e = 0.05$ and $\sigma_i=2.5\dg$) Rayleigh parameters for their distributions.
The fourth column shows the values from Table 1 of \citet{fan13}. 

We find that under this definition, more compatible with a giant impact phase, the majority ($\approx 60-95\%$) of our sampled space is unstable for an inserted planet. Similar to the FM13-style tests, we also see that instability increases with multiplicity: an additional $\sim 20\%$ of sampled space is unstable for $N\geq5$ compared to $N=2$. Increasing the eccentricity and inclination Rayleigh parameters from $\sigma_e = 0.01$ and $\sigma_i=0.5\dg$ to $\sigma_e = 0.05$ and $\sigma_i=2.5\dg$ also results in an additional $\sim 20\%$ of unstable space between observed planet pairs.

A particular observational selection effect may be consistent with dynamical packing increasing with multiplicity. Although it is uncommon, we estimate that a non-negligible proportion ($\approx 24\%$ for $N\geq2$) of observed adjacent pairs have undetected planets located in between them (see Appendix~\ref{appendix:undetected-planets} for more details). In particular, our estimated frequency of these in-between undetected planets decreases as multiplicity increases, meaning that low multiplicity systems may be more dynamically packed than suggested by Table~\ref{tab:summary-expanded-mean}. We caution that further testing is necessary to validate our estimated frequencies and their applicability to the observed Kepler multi-planet systems, however.

\begin{figure*}
  \centering
    \includegraphics[width=0.8\textwidth]{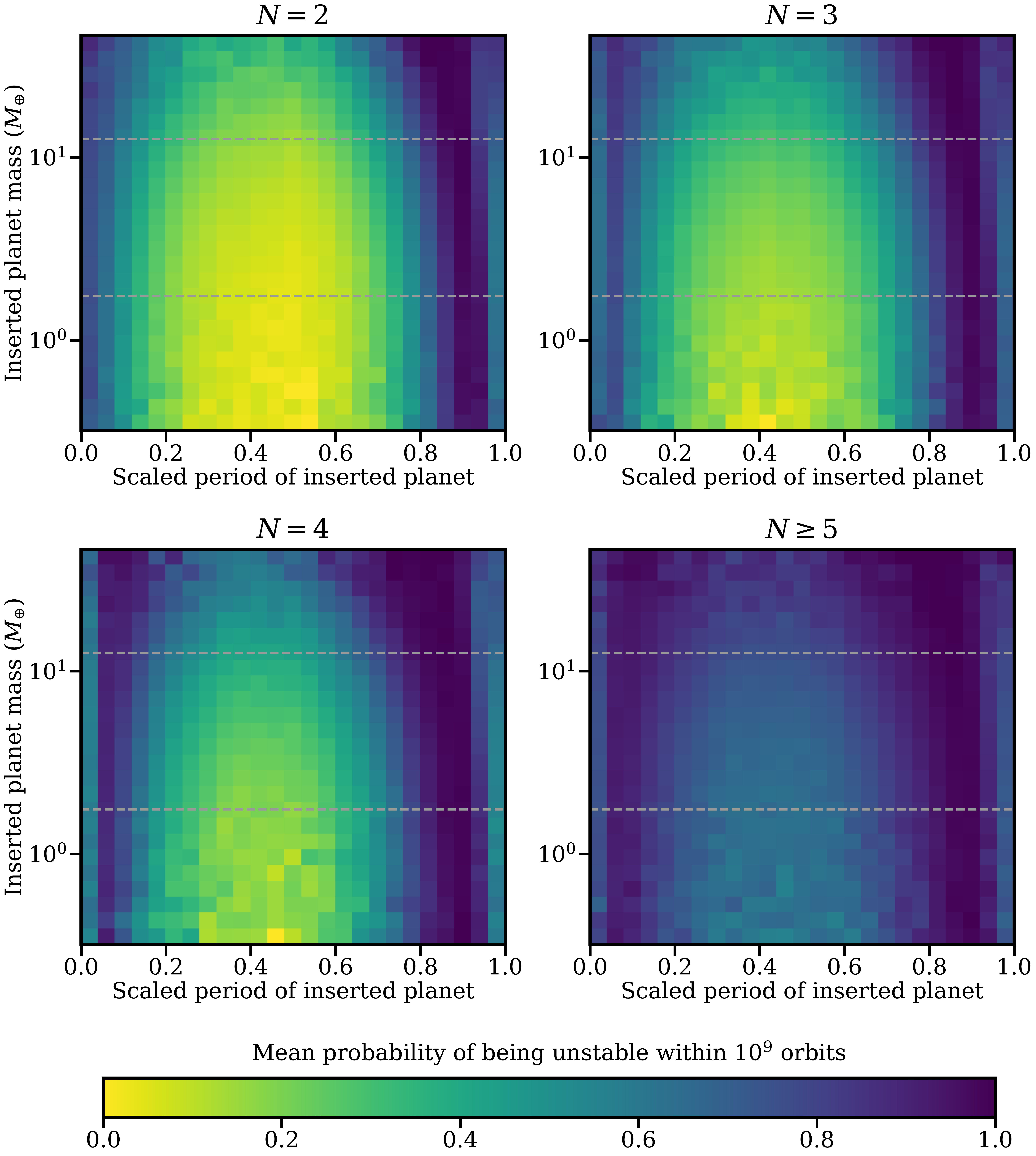}
    \caption{Similar instability ``grid plots" as Figure~\ref{fig:sample-grid-plots}, except for all sets of adjacent pairs (i.e. ``stacked grid plots") with Rayleigh distributions $\sigma_e = 0.01$, $\sigma_i = 0.5\dg$. Periods have been linearly scaled with the inner planet at 0 and the outer planet at 1. Each subplot shows probabilities according to observed multiplicities of the adjacent pairs' systems. The inserted planets' scaled periods are distributed uniformly and the masses according to Figure~\ref{fig:inserted-planet-mass-hist} (the vertical limits correspond to 95\% of masses and 68\% of masses fall within the region marked by the horizontal dashed lines).}
    \label{fig:stacked-grid-plots}
\end{figure*}

We also visually examine the proportion of unstable space for tests using Rayleigh parameters $\sigma_e = 0.01$ and $\sigma_i=0.5\dg$ in Fig.~\ref{fig:stacked-grid-plots}. These are similar to the specific examples in Fig.~\ref{fig:sample-grid-plots}, except we overlay the results for all adjacent pairs in our sample. Since each pair has its inner and outer planet at different periods, the period of the inserted planet has been scaled linearly so that the inner planet is at 0 and the outer planet is at 1 (i.e., $P_{scaled}$ = $(P_{inserted} - P_1) / (P_2 - P_1)$).

The trend of increasing instability with multiplicity is evident. We also see, unsurprisingly, that instability increases with the mass of the inserted planet (the vertical scale is logarithmic however, so the trend with mass is not very strong) and its proximity to either the inner or outer observed planet (with the exception of the ``co-orbital" columns, as discussed in Sec.~\ref{sec:results-expanded} and Appendix~\ref{appendix:co-orbitals}).

Although Fig.~\ref{fig:stacked-grid-plots} shows (qualitatively) that there is a significant amount of stable space for an additional planet in $N=2$ systems (and that the amount of space decreases with multiplicity), the majority of our sampled space between adjacent pairs is unstable according to Table~\ref{tab:summary-expanded-mean}. In other words, under the ergodic hypothesis and using our chosen distributions, observed adjacent pairs are typically strongly dynamically packed.

\subsubsection{Dynamical Packing Under the Ordered Hypothesis of Planet Formation}
\label{sec:results-ordered}

\begin{table}
\centering
\caption{Proportions of unstable configurations for the expanded parameter space tests applied to the Kepler-GAIA catalogue. Column 1 shows the host system's observed planet multiplicity, with the number of analysed adjacent pairs in parentheses. Columns 2 and 3 show the mean probability (using each pair's \textbf{most stable configuration}) for the tests with ($\sigma_e, \sigma_i$) = ($0.01, 0.5\dg$) and ($0.05, 2.5\dg$), respectively. Column 4 is from Table~1 of \citet{fan13}. Errors are calculated as in Table \ref{tab:summary-fm13-style-q16}.}
\label{tab:summary-expanded-msc}
\begin{tabular}{cccc}
\hline
Multiplicity & $\sigma_e=0.01$, $\sigma_i=0.5\dg$ & $\sigma_e=0.05$, $\sigma_i=2.5\dg$ & FM13 \\
\hline
N = 2 (408) & 22.7 $\pm$ 1.6\% & 47.7 $\pm$ 2.1\% & $\geq$31\% \\ 
N = 3 (282) & 31.4 $\pm$ 1.9\% & 60.4 $\pm$ 2.3\% & $\geq$35\% \\
N = 4 (144) & 32.3 $\pm$ 2.3\% & 70.5 $\pm$ 2.8\% & $\geq$45\% \\
N$\geq$ 5 (85) & 43.0 $\pm$ 3.6\% &  83.6 $\pm$ 2.9\% & \\
\hline
\end{tabular}
\end{table}

An opposite end-member scenario compared to the ergodic hypothesis, the ordered hypothesis assumes planets form in their most stable configuration (MSC). Therefore, we assess dynamical packing using the probability $p_{MSC}$ of the most stable configuration.
We discuss how we select this most stable configuration from each adjacent pair's probability distribution in Appendix~\ref{appendix:msc} to be more robust against outlier probabilities assigned by SPOCK.

Table~\ref{tab:summary-expanded-msc} reports the estimated fractions (with statistical uncertainties) of adjacent pairs that are unstable even when the additional planet is inserted in its most-stable configuration.
Analogous to Table~\ref{tab:summary-expanded-mean}, the central columns explore the dependence on the assumed distributions of eccentricities and inclinations.
Similar to all of our previous tests, we see that the fraction of unstable pairs increases with multiplicity: under this definition, roughly twice as many systems with $N \geq 5$ observed planets are maximally packed as those with $N=2$. We note that while our selection of the most stable configuration introduces uncertainty in our quoted packing fractions (which we estimate are $\sim 10\%$, see Appendix~\ref{appendix:msc}), the trend with multiplicity is robust regardless of how we select the MSC.
Additionally, increasing the eccentricity and inclination Rayleigh parameters from $\sigma_e = 0.01$ and $\sigma_i=0.5\dg$ to $\sigma_e = 0.05$ and $\sigma_i=2.5\dg$ approximately doubles the proportion of packed systems for all multiplicities. 

Even under this most optimistic scenario for stability, there are still a substantial number of cases where an additional planet is not stable when inserted between an observed adjacent pair.
These unstable fractions are lower than their counterparts under the ergodic hypothesis in Table~\ref{tab:summary-expanded-mean}, but provide useful lower limits for the frequency of maximally packed planet pairs.

If one wanted to use the most stable configuration to determine these lower limits for a population of multi-planet systems, it is useful to know if performing stability tests across a vast parameter space is necessary or if this can be achieved using point estimates for the MSC (e.g. if CPU hours are limited). For this, we can compare our FM13-style results (column 4 of Table~\ref{tab:summary-fm13-style-keplergaia}) to our expanded parameter space results using our lowest eccentricity and inclination Rayleigh parameters (column 2 of Table~\ref{tab:summary-expanded-msc}). We see that the proportions of unstable configurations are similar across multiplicities. This similarity is predominantly because the scenarios shown in the upper left and lower right panels of Fig.~\ref{fig:sample-grid-plots} are typical, whereas the scenarios shown in the upper right and lower left panels are not.

In other words, placing the inserted planet equidistant (in mutual Hill radii) between an adjacent pair is a good approximation of the most stable configuration when considering the lower limits of dynamical packing for a sample of multi-planet systems. There are likely other approximations which may capture different dynamical arguments for high stability (e.g. the underlying physics of the brightest yellow column shown in the lower left panel of Fig.~\ref{fig:sample-grid-plots}) or which use a different approach to determining the most stable configuration than ours (see Appendix~\ref{appendix:msc}), but this is well beyond the scope of our study.

\section{Conclusions}
\label{sec:conclusions}

In this work, we have re-examined the dynamical packing of the Kepler multi-planet systems through a new lens.
Previous work has focused on dynamical packing as a binary property: the yes/no question of whether, given an adjacent pair of observed planets, at least one possible configuration exists for an additional planet between them that could survive over long timescales.
This theoretical, binary question of whether such stable arrangements exist is only meaningful to the extent that the planet formation process is capable of generating such configurations, however.
We therefore argue for a more meaningful, explicit definition of dynamical packing that is instead continuous: for a given pair of observed, adjacent planets, what is the probability that the planet formation process could have created an additional planet between them that would have survived to the present day?

The obvious complication is that this new definition requires a detailed planet formation model that provides a probability distribution for new planets over all possible configurations. Since planet formation modelling is still an open area of research, choosing a detailed planet formation model is not a straightforward task. Instead, we simplify matters by considering two frameworks of planet formation.

One possibility is that the formation process is sufficiently dissipative and gradual that planets would naturally fall into the lowest-energy, most stable configurations available (which we refer to as the ``ordered hypothesis").
In this limit, our continuous definition reduces to the simple binary classification considered in previous studies. Perhaps for that reason, as well as the enormous computational benefit of only testing a single point estimate of the most stable configuration, this \citet{fan13} used this as their measure for the dynamical packing of the Kepler multi-planet systems.
It is not clear that this framework of planet formation occurs, or that the system remains in this state, however. For example, one scenario where planets form in their most stable configuration is settling into resonant chains. 
Such lowest-energy resonant chains are rarely observed for Kepler multi-planet systems, though\citep{fab14}. Indeed, several studies have argued that if planets settle into resonant chains during the dissipative gas disk phase, then chaotic dynamics must subsequently destabilize these resonant chains following disk dissipation \citep[e.g.][]{izi17,mat20,pic20,izi21,est22,gol22}. If so, chaotic dynamics could also remove planets from a different type of most stable configuration.

As long as we broaden `planet formation' to include this violent final stage of planetary growth through collisions (the so-called giant impacts phase), we would expect planet formation to rarely populate small pockets of stability in a vast phase space.
Unlike an ordered picture of planet formation in which the existence of small stability pockets matters, the relevant metric in this chaotic picture of planet formation (including giant impacts) is the fraction of the vast phase space between the pair that would allow an additional planet to form and survive to the present day.
\citet{tre15} has argued that for a sufficiently chaotic giant impact phase, one would expect planet formation to approximately fill the phase space of orbital configurations uniformly.
Subsequently, the distribution of orbital architectures for observed systems corresponds to the subset of phase space that is dynamically stable on timescales comparable to the systems' ages.
Assuming a simplified stability criterion, \citet{tre15} shows that this ``ergodic hypothesis" predicts distributions of orbital eccentricities and interplanetary spacings that are broadly consistent with observations.

Both the ordered and ergodic hypotheses are idealized pictures of planet formation and we expect that observed planetary systems have formation histories falling somewhere in between these two limits. Since these two hypotheses represent opposing scenarios, they provide intuition for the possible range of maximally packed planetary systems.

Using the ergodic definition and our chosen distributions, observed pairs of planets in compact multi-planet systems are typically strongly packed: $\sim 60-95\%$ (Table \ref{tab:summary-expanded-mean}). 

Under the more restrictive ordered definition of maximally packed, one might expect that a smaller proportion of systems would be maximally packed. This is in fact seen in a comparison between Table~\ref{tab:summary-expanded-mean} and Table~\ref{tab:summary-expanded-msc}. Nevertheless, our results show that a significant portion of adjacent pairs, $\sim 20-80\%$, are maximally packed using the more stringent definition.

Regardless of our definition of dynamical packing, we see a clear trend of increasing packing with increasing multiplicity, assuming that eccentricities are similar across multiplicities. This is consistent with the observation that the interplanetary spacing (as measured by e.g. period ratio or Hill-radius separation) between observed adjacent pairs decreases as the multiplicity increases \citep[e.g.][]{wei18,zhu21}.
As the separation between a pair decreases, more mean motion resonances can overlap to drive chaos and instabilities in multi-planet systems \citep{dec13}, which may be responsible for interplanetary spacing constraints in very compact multi-planet systems \citep{obe17}.

One possible source of this trend of increased packing with multiplicity is an observational selection effect: systems with low observed multiplicity may be more likely to have undetected planets located between observed adjacent pairs. When using \texttt{SysSimExClusters} \citep{he19} to generate a set of ``Kepler-like" multi-planet systems and to simulate observing them with the Kepler telescope and its detection pipeline, we saw that lower multiplicity systems had a higher proportion of in-between undetected planets (see Appendix~\ref{appendix:undetected-planets} for more details). In other words, low multiplicity systems could be more dynamically packed than they appear.

As noted in Sec.~\ref{sec:results-expanded}, observed adjacent pairs with low $p_{mean}$ (i.e. their grid plots, like those in Fig.~\ref{fig:sample-grid-plots}, look very yellow) are suitable candidates to search for these in-between undetected planets. Grid plots and summary statistics for all observed adjacent pairs in our Kepler-GAIA catalogue are openly available online.

While such a trend with multiplicity could simply be an imprint of formation in the protoplanetary disk, it could also naturally arise through later dynamical instabilities causing giant impacts.
A giant impact phase, likely with a tail of collisions spanning the Gyr lifetimes of typical Kepler stars, would generate larger interplanetary gaps as planets merge. This process results in progressively longer-lived, lower-multiplicity systems.
Though we have not specifically investigated the question here, hypotheses involving this sculpting of multi-planet systems \citep{pu15, volk15} might naturally predict more space for additional planets (i.e., lower dynamical packing) for low-multiplicity systems that have had several planets removed.

Alternatively, our assumption of a constant eccentricity distribution with multiplicity may not hold.
There is some observational evidence suggesting that systems with $N=2$ have higher eccentricities than those with $N\geq3$ \citep{xie16}. Detailed modelling of Kepler multi-planet system architectures similarly shows higher eccentricity in low multiplicity systems \citep{he20}. Furthermore, observational modelling suggests that orbital inclinations decrease significantly with increasing multiplicity \citep{zhu18a,mil21}.
Given that eccentricities are comparable to inclinations in most astrophysical disks (and indeed, this was seen in the modelling by \citet{he20}), a decreasing trend of eccentricity with inclination therefore seems plausible.
While we do not consider systems with a single transiting planet, eccentricities extracted from transit durations in single-transiting systems are roughly four times higher than those in multi-planet systems \citep{mil19,eyl19}, consistent with the above picture.

Comparing our $N=2$ results using a higher eccentricity Rayleigh parameter to our $N\geq5$ results using a lower eccentricity parameter in Tables~\ref{tab:summary-expanded-mean} and~\ref{tab:summary-expanded-msc}, we see similar values. It may in fact be that systems with different multiplicities have comparable packing which could be caused by the same processes generating the eccentricity differences.
\cite{daw16} have argued that a giant impact phase might establish an equilibrium between scatterings that excite eccentricities and collisions that dampen them.

We expect that the relationships between the physical processes of planet formation and evolution (including giant impacts), present-day eccentricities in mature systems, the degeneracy between observed vs. intrinsic multiplicity and mutual inclinations (including in-between undetected planets), where observed systems fall between the ordered and ergodic hypotheses, and dynamical packing are complex and very intertwined. This paper is a first attempt at examining dynamical packing in an expanded context and searching for clues to help disentangle these relationships. In the past, this was not feasible due to practical constraints: limited computation time for running Nbody integrations of the various configurations needed to explore such a large parameter space. With \spock, that constraint is eased substantially and opens up the option, as we explored, of using a broader definition of dynamical packing that's physically tied to specific models of planet formation. We certainly do not expect that the methods and definitions we utilized are perfect, but they have illuminated a strong need for the exoplanet community to come together and consider dynamical packing explicitly linked to planet formation. 

\section*{Acknowledgements}

We are grateful for Gwendolyn Eadie's and Joshua Speagle's expertise, insight, and patience in discussions of our error estimates. We thank Hanno Rein and Kristen Menou for their thoughtful discussion on several occasions as we worked on this project. Additionally, we thank the anonymous reviewer for their extraordinarily valuable comments, questions, and suggestions.

AO was partially supported by the Natural Sciences and Engineering Research Council of Canada (NSERC) during this project (application ID 504349-2017). NM and AO were partially supported by NSERC during this project (application ID RGPIN-2017-06459)

This research has made use of the NASA Exoplanet Archive, which is operated by the California Institute of Technology, under contract with the National Aeronautics and Space Administration under the Exoplanet Exploration Program.
This research has made extensive use of the pandas\citep{pandas}, NumPy\citep{numpy}, Astropy\footnote{\url{http://www.astropy.org}} \citep{astropy:2013, astropy:2018, astropy:2022} and Matplotlib\citep{matplotlib} packages for python\footnote{\url{https://www.python.org/}}.

The University of Toronto operates on the traditional land of the Huron-Wendat, the Seneca, and the Mississaugas of the Credit. Harvey Mudd College operates on Torojoatngna, one of many villages in the traditional lands of the Tongva-Gabrielino/Gabrieleño/Kizh peoples. We encourage readers to learn whose lands they occupy and how to support these Indigenous Peoples.

\section*{Data Availability}

We obtained the Kepler quarters 1--6 data \citep{bat13} used in this paper from the files \texttt{table3.dat, table4.dat,} and \texttt{table5.dat} located at \url{http://cdsarc.unistra.fr/viz-bin/cat/J/ApJS/204/24}.

We obtained the cumulative Kepler data from the NASA Exoplanet Archive\citep{cumulative}, which we accessed on 2020-09-21 at 15:02 EDT. After applying the two filters described in Sec.~\ref{sec:methods-keplergaia}, 4612 rows were returned. Details about the cumulative catalogue's generation are available at \url{https://exoplanetarchive.ipac.caltech.edu/docs/PurposeOfKOITable.html}

We obtained the revised Kepler/Gaia stellar parameters \citep{ber20a} from the file \texttt{table2.dat} located at \url{https://cdsarc.cds.unistra.fr/viz-bin/cat/J/AJ/159/280} and the planetary parameters \citep{ber20b} from the file \texttt{table1.dat} located at \url{https://cdsarc.cds.unistra.fr/viz-bin/cat/J/AJ/160/108}.

Our code used to combine the various catalogues, generate and integrate our \rebound~simulations, and obtain \spock probabilities is available at \url{github.com/aobertas/dynamical-packing-kepler-multis}. Summary metrics and grid plots (similar to Fig.~\ref{fig:sample-grid-plots}) for all observed adjacent pairs in our Kepler-GAIA catalogue are also available online via the GitHub repository.


\bibliographystyle{mnras}
\bibliography{bibliography}



\appendix

\section{Equidistant mutual hill radius}
\label{appendix:mutual-hill-radius}

For calculating the inserted planet's placement, we defined the mutual Hill radius of planets $i$ and $j$ (with masses $M_i$ and $M_j$ and semimajor axes $a_i$ and $a_j$ orbiting a star with mass $M_{\star}$) as

\begin{equation}
\label{eq:mutual-hill-radius}
    R_{\mathrm{Hill},i,j} = \left(\frac{M_i + M_j}{3M_{\star}}\right)^{1/3}\left(\frac{a_i + a_j}{2}\right)
\end{equation}

The semimajor axis of the inserted planet is then,

\begin{eqnarray}
    a_{\mathrm{insert}} &= a_{\mathrm{inner}} + \Delta R_{\mathrm{Hill},\mathrm{insert},\mathrm{inner}} \\
    &= a_{\mathrm{outer}} - \Delta R_{\mathrm{Hill},\mathrm{insert},\mathrm{outer}}
\end{eqnarray}

The spacing $\Delta$ can then be solved for in terms of the masses of the three planets and the two known semimajor axes. Defining the parameter $X_{i,j}$ as,

\begin{equation}
    X_{i, j} = \frac{1}{2} \left(\frac{m_i + m_j}{3M_{\star}}\right)^{1/3}
\end{equation}

This gives a solution\footnote{This involves solving a quadratic, but the other solution would place the inserted planet far outside of the outer planet.} for the spacing $\Delta$ as

\begin{equation}
    \Delta = \frac{(X_{\mathrm{i}} + X_{\mathrm{o}}) (a_{\mathrm{i}} + a_{\mathrm{o}}) - \sqrt{(X_{\mathrm{i}} + X_{\mathrm{o}})^2 (a_{\mathrm{i}} + a_{\mathrm{o}})^2 - 4 X_{\mathrm{i}} X_{\mathrm{o}} (a_{\mathrm{o}} - a_{\mathrm{i}})^2}}{2 X_{\mathrm{i}} X_{\mathrm{o}} (a_{\mathrm{o}} - a_{\mathrm{i}})}
\end{equation}

Where $a_{\mathrm{inner}}$, $a_{\mathrm{outer}}$, $X_{\mathrm{insert,inner}}$, $X_{\mathrm{insert,outer}}$ have been written as $a_{\mathrm{i}}$, $a_{\mathrm{o}}$, $X_{\mathrm{i}}$, and $X_{\mathrm{o}}$ respectively to condense the expression.

\section{Estimating frequency of undetected planets located between observed adjacent pairs}
\label{appendix:undetected-planets}

While numerous studies have examined the intrinsic multiplicity distribution of Kepler systems \citep[e.g.][]{tra16, zhu18a, zin19, san19}, these do not report metrics related to the locations of undetected (i.e. non-transiting) planets in relation to observed (i.e. transiting) planets. We perform a simple test to estimate the frequency of undetected planets being located between observed adjacent pairs.

We began by generating a ``simulated" catalogue of Kepler-like systems using the \texttt{SysSimExClusters} code\footnote{We used a version from October 2019 for our analysis (GitHub commit ``a0b88bbfa2ec95a5d64382b102df1f0f32cd66be").} developed by \citet{he19} and their reported best-fit parameters for the ``Clustered Periods and Sizes" model. Next, we generated an ``observed" catalogue by running the simulated catalogue through their Kepler observation pipeline code. Finally, we compared the two catalogues in order to estimate the frequency of detected adjacent pairs with at least one undetected planet in between them. The simulated catalogue has 45 652 systems with 200 034 planets. There are 1624 systems and 2151 planets in the observed catalogue. Those same observed catalogue systems have 8844 planets in the simulated catalogue (i.e. there are 6693 undetected planets).

To estimate this frequency, we considered all systems in the observed catalogue for which each individual system had at least $N$ detected (i.e. transiting) planets (with a minimum of 2 so there is at least one detected adjacent pair), with the additional criterion of having an orbital period $P < 100$ days to be classified as detected. Next, we compared each system's observed catalogue planets to its corresponding simulated catalogue planets to determine the relative locations of each system's detected and undetected planets, according to their orbital periods. For each system, we counted the number of detected planets, undetected planets (note that these could be located interior to, in between, or exterior to detected planets), detected adjacent pairs, and the number of detected pairs with at least one undetected planet located in between (i.e. a detected adjacent pair with three undetected planets in between is counted the same as a pair with only one undetected planet). Table~\ref{tab:undetected-planets} reports the sums of these values across systems, for different values of the systems' observed multiplicities $N$. Additionally, the last column shows the corresponding fraction of detected pairs with undetected planets in between $f_{p,u}$ (i.e. the estimated frequency).

The estimated frequency that a detected adjacent pair has at least one undetected planet is $f_{p,u} \approx 24\%$ ($\sim$ 1 in 4 pairs). In other words, there is a non-negligible likelihood that a detected adjacent pair in our Kepler-GAIA catalogue (Sec.~\ref{sec:methods-keplergaia}) is not truly adjacent. Furthermore, higher observed multiplicity systems have smaller estimated frequencies than lower multiplicity systems (i.e. the likelihood of being truly adjacent is smaller for $N=3$ and even smaller for $N=2$ systems).

We caution that our analysis is a simple test in order to estimate this frequency for the purpose of interpreting our results in Sec.~\ref{sec:results-expanded}. A more thorough and robust analysis is necessary to validate the frequencies shown in Tables~\ref{tab:undetected-planets} for the observed Kepler multi-planet systems.

\begin{table}
\centering
\caption{Counts of the number of observed systems $n_s$, the number of detected (i.e. transiting with $P < 100$ days) planets $n_d$, the number of undetected planets $n_u$, the number of adjacent detected pairs $n_p$, the number of pairs with at least one undetected planet in between them $n_{p,u}$, and the fraction of pairs with undetected planets $f_{p,u} = n_{p,u} / n_p $ (i.e. the estimated frequency) for our simulated and observed catalogues generated by \texttt{SysSimExClusters}. The counts of $n_d$, $n_u$, $n_p$, and $n_{p,u}$ are summed across all $n_s$ systems. Each row shows values for systems with observed planet multiplicity $N$.}
\label{tab:undetected-planets}
\begin{tabular}{ccccccc}
\hline
Multiplicity & $n_{s}$ & $n_{d}$  & $n_{u}$ & $n_{p}$ & $n_{p,u}$ & $f_{p,u}$\\
\hline
N$\geq$ 2 & 350 & 834 & 1205 & 484 & 114 & 0.2355 \\
N = 2 & 250 & 500 & 901 & 250 & 76 & 0.3040 \\ 
N = 3 & 74 & 222 & 238 & 148 & 29 & 0.1959 \\
N = 4 & 20 & 80 & 54 & 60 & 7 & 0.1167 \\
N$\geq$ 5 & 6 & 32 & 12 & 26 & 2 & 0.0769 \\
\hline
\end{tabular}
\end{table}

\section{Reliability of SPOCK}

\subsection{Comparison Between Nbody Integrations and \spock}
\label{appendix:nbody-spock comparison}

Although \spock has been thoroughly tested \citep{tam20,tam21b}, we present a comparison of Nbody integrations and \spock for our FM13-style tests using the Kepler-GAIA catalogue (see Sec.~\ref{sec:methods-fm13style}) to check its performance for our specific use.

\begin{figure}
  \centering
    \includegraphics[width=\columnwidth]{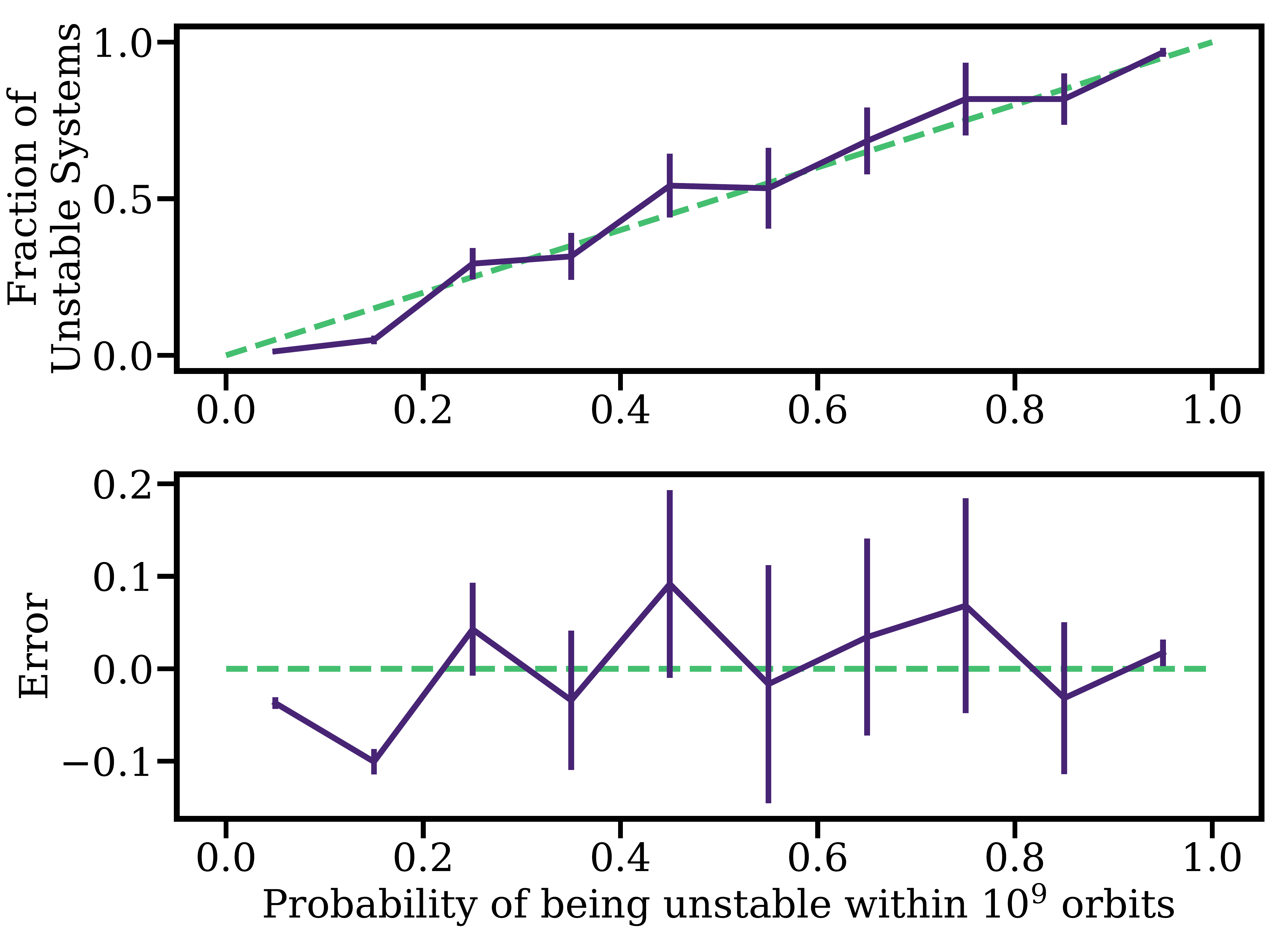}
    \caption{Top: The solid purple line shows the fraction of adacent pairs in the FM13-style test using the Kepler-GAIA catalogue whose Nbody integrations go unstable within $10^9$ orbits after inserting an additional planet, binned by their corresponding probabilities. The dashed green line shows a 1:1 line (i.e. if \spock had perfect performance, the solid purple line would lay directly on top of the dashed green line). Bottom: The difference between the top panel's solid purple line and the dashed green line. Each bin's error estimate is binomial.}
    \label{fig:spock-performance-binned}
\end{figure}

Fig.~\ref{fig:spock-performance-binned} shows this performance. In the top panel, we binned each tested adjacent pair based on its probability of being unstable (from \spock), then calculated the fraction of Nbody integrations which had an instability before the maximum integration time ($10^9$ orbits) for each bin (shown as the solid purple line). Each bin has a binomial error. The dashed green line shows a 1:1 relationship (i.e. if \spock had perfect performance, the purple line would lay directly on top of the green line). In the bottom panel, we show the difference between the top panel's purple and green lines. We see that \spock performs well for the ensemble of adjacent pairs.

There is some discrepancy between Nbody integrations and \spock when considering adjacent pairs based on their systems' observed multiplicities, however. Columns 3 and 4 of Table~\ref{tab:summary-fm13-style-keplergaia} show the mean probability of being unstable for $10^9$ orbits based on Nbody integrations and \spock. While the values for $N=3$ and $N=4$ are quite similar, the values $N=2$ and $N\geq5$ are not. The two mean probabilities for $N\geq$ are within errors, but this is not the case for $N=2$. The primary cause of this difference is that the probability of being unstable is non-zero, even if it is small. The mean probability for Nbody integrations is calculated by taking the average of 1's and 0's. The mean probability for \spock values is calculated by taking the average of (mostly) small numbers close to 0 (but never 0) and larger numbers close to 1 (which can include 1\footnote{This will happen if a system has an instability within \spock's $10^4$ orbit integration time}), however. In other words, the mean probability of being unstable using \spock values can be biased when the sample contains a large proportion of stable systems.

If we were to assume the same differences between Columns 3 and 4 in Table~\ref{tab:summary-fm13-style-keplergaia} for Columns 2 in Table~\ref{tab:summary-expanded-msc} (which only uses \spock values to calculate mean probabilities), then a smaller proportion of pairs in $N=2$ systems and a larger proportion of pairs in $N\geq5$ systems would be unstable (although perhaps less so, due to the difference being within errors). This would strengthen the trend of increasing dynamical packing with increasing multiplicity.

\subsection{Co-orbital Configurations}
\label{appendix:co-orbitals}

Some configurations where the inserted planet is effectively ``co-orbital" with either the inner or outer observed planet (visible in the first and last columns of the grid plots in Figs.~\ref{fig:sample-grid-plots} and~\ref{fig:stacked-grid-plots}) show low instability which is likely an artefact of \spock's operating procedures.

Essentially, \spock integrates a system for $10^4$ orbits and monitors a set of parameters, which are physically relevant to the dynamics at play (see \citet{tam20} for a full and detailed description of \spock's methodology). \spock was trained on how these parameters vary throughout the integration.

Planets with these co-orbital configurations have synodic periods comparable to \spock's integration time. At most, there may be only a few conjunctions. These configurations were not included in \spock's training set \citep{tam20}, so \spock's probability does not appropriately incorporate the signature of their dynamical interactions on the set of relevant parameters. The probabilities for these co-orbital configurations are therefore unreliable.

We opted to retain these configurations for all quantities computed and presented in this paper, primarily to avoid imposing restrictions on the data. As a check of their effect, removing configurations laying in the first and last column of the grid plots (Fig.~\ref{fig:sample-grid-plots}) typically changes all numbers Tables~\ref{tab:summary-expanded-mean}, ~\ref{tab:summary-expanded-msc}, ~\ref{tab:summary-expanded-msc-1}, and~\ref{tab:summary-expanded-msc-10} by quantities less than error. Table~\ref{tab:summary-expanded-msc-minimum} is the exception, although changes are typically within twice the error.

We note that probabilities for configurations not included in \spock's training set are not inherently unreliable, however. For example, \spock's training set only included 3-planet systems but \spock is (in general) reliable for systems of higher multiplicities. Additional planets add extra sources for gravitational perturbations that can lead to instability, but the signatures of these interactions on the set of relevant parameters are (in general) similar to those in 3-planet systems.

\section{Determining the Most Stable Configuration}
\label{appendix:msc}

In Sec.~\ref{sec:results-ordered} (including Table~\ref{tab:summary-expanded-msc}), we quantify the proportion of dynamically packed adjacent pairs using each pair's most stable configuration (MSC) probability. Here, we describe how we defined this probability.

While there are many ways to approach this task, we chose to consider several percentiles of each pair's probability distribution (note: each pair has 5000 probabilities corresponding to its sampled configurations with an inserted planet). This is a straightforward computation that allows a highly stable configuration to be determined without any complications or caveats from user-imposed restrictions or requirements (e.g. such as by making phase space cuts that are applied differently to each pair based on physical arguments).

In principle, the minimum probability is the most stable configuration and we show the mean probabilities using this in Table~\ref{tab:summary-expanded-msc-minimum}. In practice, the minimum probability could be an extreme outlier that misrepresents the low-probability tail of the distribution.

One source of these outliers are low-value but unreliable \spock~probabilities. For example, configurations where the inserted planet is effectively ``co-orbital" with either the inner or outer observed planet (visible in the first and last columns of the grid plots in Figs.~\ref{fig:sample-grid-plots} and~\ref{fig:stacked-grid-plots}. We discuss these configurations in more detail in Appendix~\ref{appendix:co-orbitals}), but the metric most impacted is the minimum probability.

Another source of outliers are configurations which truly and reliably have low probabilities, but are located in a region of phase space which is overwhelmingly unstable on average. For example, the upper right panel of Fig.~\ref{fig:sample-grid-plots} shows an adjacent pair whose FM13-style configuration (denoted by the white X) has a low probability. It is situated in a region which typically has high instability probabilities, however (i.e. it lays on a very purple spot as opposed to a very yellow spot). 

Rather than using the minimum probability, we consider three different percentiles as candidates for a more robust MSC. Table~\ref{tab:summary-expanded-msc-1} shows mean instability probabilities using each pair's first percentile probability as its MSC. Table~\ref{tab:summary-expanded-msc-5} shows the same, but with the fifth percentile (note that this is identical to Table~\ref{tab:summary-expanded-msc}) and Table~\ref{tab:summary-expanded-msc-10} with the tenth percentile.

The mean probabilities depending on the percentile chosen and the range is on the order of 10 percentage points. This is similar to the largest discrepancy we saw between the Nbody integration and \spock{} for the FM13-style tests (columns 3 and 4 of Table~\ref{tab:summary-fm13-style-keplergaia}. It's reasonable then to say that we can't determine the proportion of dynamically packed systems with greater precision than $\sim10\%$. Nevertheless, whichever choice we make for the percentile, we see that dynamical packing increases with multiplicity, so this trend is robust.

\begin{table}
\centering
\caption{Proportions of unstable configurations for the expanded parameter space tests applied to the Kepler-GAIA catalogue. Column 1 shows the host system's observed planet multiplicity, with the number of analysed adjacent pairs in parentheses. Columns 2 and 3 show the mean probability (using each pairs's \textbf{minimum probability}) for the tests with ($\sigma_e, \sigma_i$) = ($0.01, 0.5\dg$) and ($0.05, 2.5\dg$), respectively. Column 4 is from Table~1 of \citet{fan13}. Errors are calculated as in Table \ref{tab:summary-fm13-style-q16}.}
\label{tab:summary-expanded-msc-minimum}
\begin{tabular}{cccc}
\hline
Multiplicity & $\sigma_e=0.01$, $\sigma_i=0.5\dg$ & $\sigma_e=0.05$, $\sigma_i=2.5\dg$ & FM13 \\
\hline
N = 2 (408) & 8.1 $\pm$ 1.0\% & 18.7 $\pm$ 1.5\% & $\geq$31\% \\ 
N = 3 (282) & 10.8 $\pm$ 1.0\% & 24.4 $\pm$ 1.8\% & $\geq$35\% \\
N = 4 (144) & 10.9 $\pm$ 1.3\% & 25.2 $\pm$ 2.1\% & $\geq$45\% \\
N$\geq$ 5 (85) & 14.9 $\pm$ 2.1\% &  39.2 $\pm$ 3.5\% & \\
\hline
\end{tabular}
\end{table}

\begin{table}
\centering
\caption{Same as Table~\ref{tab:summary-expanded-msc-minimum}, except using the \textbf{1st-percentile probability} for each adjacent pair to calculate mean values.}
\label{tab:summary-expanded-msc-1}
\begin{tabular}{cccc}
\hline
Multiplicity & $\sigma_e=0.01$, $\sigma_i=0.5\dg$ & $\sigma_e=0.05$, $\sigma_i=2.5\dg$ & FM13 \\
\hline
N = 2 (408) & 16.0 $\pm$ 1.4\% & 37.1 $\pm$ 2.0\% & $\geq$31\% \\ 
N = 3 (282) & 21.5 $\pm$ 1.6\% & 48.1 $\pm$ 2.3\% & $\geq$35\% \\
N = 4 (144) & 20.5 $\pm$ 1.8\% & 55.9 $\pm$ 2.9\% & $\geq$45\% \\
N$\geq$ 5 (85) & 28.7 $\pm$ 3.0\% &  69.8 $\pm$ 3.5\% & \\
\hline
\end{tabular}
\end{table}

\begin{table}
\centering
\caption{Same as Table~\ref{tab:summary-expanded-msc-minimum}, except using the \textbf{5th-percentile probability} for each adjacent pair to calculate mean values. Identical to Table~\ref{tab:summary-expanded-msc}, but repeated for ease of comparison.}
\label{tab:summary-expanded-msc-5}
\begin{tabular}{cccc}
\hline
Multiplicity & $\sigma_e=0.01$, $\sigma_i=0.5\dg$ & $\sigma_e=0.05$, $\sigma_i=2.5\dg$ & FM13 \\
\hline
N = 2 (408) & 22.7 $\pm$ 1.6\% & 47.7 $\pm$ 2.1\% & $\geq$31\% \\ 
N = 3 (282) & 31.4 $\pm$ 1.9\% & 60.4 $\pm$ 2.3\% & $\geq$35\% \\
N = 4 (144) & 32.3 $\pm$ 2.3\% & 70.5 $\pm$ 2.8\% & $\geq$45\% \\
N$\geq$ 5 (85) & 43.0 $\pm$ 3.6\% &  83.6 $\pm$ 2.9\% & \\
\hline
\end{tabular}
\end{table}

\begin{table}
\centering
\caption{Same as Table~\ref{tab:summary-expanded-msc-minimum}, except using the \textbf{10th-percentile probability} for each adjacent pair to calculate mean values.}
\label{tab:summary-expanded-msc-10}
\begin{tabular}{cccc}
\hline
Multiplicity & $\sigma_e=0.01$, $\sigma_i=0.5\dg$ & $\sigma_e=0.05$, $\sigma_i=2.5\dg$ & FM13 \\
\hline
N = 2 (408) & 27.2 $\pm$ 1.7\% & 54.0 $\pm$ 2.1\% & $\geq$31\% \\ 
N = 3 (282) & 37.2 $\pm$ 2.0\% & 66.7 $\pm$ 2.2\% & $\geq$35\% \\
N = 4 (144) & 40.8 $\pm$ 2.6\% & 78.3 $\pm$ 2.5\% & $\geq$45\% \\
N$\geq$ 5 (85) & 52.2 $\pm$ 3.7\% &  88.2 $\pm$ 2.7\% & \\
\hline
\end{tabular}
\end{table}


\bsp	
\label{lastpage}
\end{document}